\documentclass[aps,prd,reprint,groupedaddress,nofootinbib,a4paper, 11pt]{revtex4}
  \usepackage{geometry}
\usepackage{hyperref}
\usepackage{tikz, amsmath} 
\usetikzlibrary{calc}
\usepackage{bbm}
\usepackage{xcolor} 
\usepackage{dsfont}
\usepackage{subfigure}
\usepackage{hyperref}
\usepackage{microtype}
\usepackage{caption}
\usepackage{amsmath} 
\usepackage{float}
\usepackage{graphics}
\usepackage{psfrag}
\usepackage{color}
\usepackage{slashed}
\usepackage{subfigure}
\usepackage{graphicx}
\usepackage{array,multirow}
\usepackage[normalem]{ulem}
\usepackage{amsfonts}
\usepackage{amssymb}
\usepackage{subfigure}
\usepackage{multirow}
\usepackage{mathtools}

\usepackage{dcolumn}
\usepackage{bm}
\usepackage{slashed}

\usepackage{epsfig}

\newcommand{\beq}{\begin{eqnarray}}
\newcommand{\eeq}{\end{eqnarray}}

\newcommand{\bmp}{\noindent\begin{minipage}{16cm}}
\newcommand{\emp}{\end{minipage}\vskip 7mm} 

\begin{document}

\title{Recursive relations from diffeomorphism in the Randall-Sundrum model}

\author{Haiying Cai$^{1}$}
\email{hcai@korea.ac.kr}
\affiliation{College of Physics and Communication Electronics, Jiangxi Normal University, Nanchang, Jiangxi 330022, China}

\author{Giacomo Cacciapaglia}
\email{cacciapa@lpthe.jussieu.fr}
\affiliation{Laboratoire de Physique Theorique et Hautes Energies, UMR 7589, Sorbonne Universit\'e \& CNRS, 4 place Jussieu, 75252 Paris Cedex 05, France.}

\vskip 1.5mm

\begin{abstract}

Models of gravity in warped extra dimensions enjoy invariance under diffeomorphism. We   derive the nonlinear transformation rules for the metric perturbations in the unitary gauge.  As an off-shell symmetry, the main consequence of diffeomorphism  is a set of recursive relations linking consecutive orders in the field expansion of the effective Lagrangian. The physical consequences are briefly explored for the Randall-Sundrum model with hard branes.

\end{abstract}

\maketitle

Warped extra dimensional spacetime with negative curvature\cite{Randall:1999vf}, akin to Anti-de Sitter (AdS) in five dimensions (5d), offers a compelling framework for addressing long-standing puzzles in particle physics, such as the Planck hierarchy problem and the weakness of the gravity coupling\cite{Randall:1999ee}. A fascinating aspect is the conjectured correspondence between the weakly coupled 5d theory of gravity in AdS and an approximately conformal field theory (CFT) residing on the 4d boundary\cite{Maldacena:1997re, Witten:1998qj, Gubser:1998bc}. From this AdS/CFT perspective, mass scales arise as the breaking of the conformal symmetry: this is achieved by cutting off the extra dimensional space via either a hard wall (brane), as in the Randall-Sundrum model\cite{Randall:1999vf, Randall:1999ee}, or by a dynamical infra-red (IR) cut-off, as in the soft-wall model\cite{Batell:2008me, Cabrer:2009we,Gherghetta:2010he}. Both constructions have been  extensively explored in the literature,  particularly due to their profound influence on our understanding of  gravitational waves\cite{Randall:2006py}, early universe cosmology\cite{Konstandin:2011dr, Baratella:2018pxi} and dark matter.  In all these distinct phenomenological applications, the symmetries of  warped extra dimension play a pivotal role. In this paper, we aim at elucidating the concept of diffeomorphism and its consequences on relevant physical examples. For convenience, we will focus our discussion on the Randall-Sundrum (RS) model with the Goldberger-Wise (GW) stabilization mechanism\cite{Goldberger:1999uk, Goldberger:1999un}, although general conclusions  apply equally to other warped theories such as soft-wall models.

Analogously to the 4d diffeomorphism of the graviton action\cite{Hinterbichler:2011tt}, the exact 5d diffeomorphism includes a nonlinear component involving the derivative form of the coordinate shift $\xi^M$ multiplied by metric perturbations.\footnote{The parameter $\xi^M$ is considered to be of  the same order as the metric perturbations.} The nonlinear part is crucial in ensuring the full action invariance. Unlike the linearized  diffeomorphism depicted in the literature\cite{Pilo:2000et, Kogan:2001qx, Chivukula:2022kju}, the exact diffeomorphism is realized in an off-shell manner: as such, the fields are  not required to obey the equations of motion (EOMs), and the symmetry imposes no constraints on the Kaluza-Klein spectrum. Two key proofs are delivered in this letter. Firstly we  proved that  the  diffeomorphism variation of 5d Lagrangian gives rise to  a total derivative: $\delta \left( \sqrt{g} L  \right)=  \partial_M \left( \xi^M \sqrt{g} L \right)$. Then we derive the  diffeomorphism transformation for  metric perturbations without approximation in the unitary gauge,   by  vanishing  the metric entry connecting the 4d Minkowsky space to the fifth dimension in the RS model (i.e. $g_{\mu 5} = 0$).   As expected, this transformation  does not mix physical fields with different spins and contains a piece of  nonlinear variation, whose operation on the $n$-th order Lagrangian expansion is equivalent  to  the linear variation of the $(n+1)$-th order terms.  Hence our main result consists of a set of simple recursive relations valid for the bulk effective Lagrangian prior to any 5d integration: due to the nonlinearity, this off-shell symmetry connects the neighboring orders in the field expansion of the bulk Lagrangian, effectively governing the interaction structure of the theory.

\vspace{0.3cm}
We start with a brief review of  the RS model\cite{Randall:1999ee} stabilized by the GW mechanism\cite{Goldberger:1999uk}. The 5d action for the metric $g$ and the GW scalar field $\phi$ is written as $S = S_{\rm bulk} + S_{\rm branes}$:
\beq
S_{\rm bulk} =  -   \frac{1}{2 \kappa^2} \int d^5 x \ \sqrt{g} \, {\cal R} + 
 \frac{1}{2} \int d^5 x \ \sqrt{g} \ g^{IJ}  \partial_I \phi \partial_J \phi  
-  \int d^5 x \ \sqrt{g} V(\phi) \,, \label{Act}
\eeq
and
\beq
S_{\rm branes} = -  \int d^5 x \ \sum_{i} \sqrt{g_4} \lambda_{i}(\phi) \delta(y - y_i) \,, \label{Act:brane}
\eeq
where $y\equiv x^5$ is the  fifth dimension coordinate and the Lorentz indices follow this notation: capital Latin indices $(I,J,\dots) \in (\mu, 5)$  span all the dimensions, while Greek  indices $(\mu, \nu,\dots)$  are assigned to the 4d Minkowski spacetime.  Eq.~\eqref{Act} consists  of the Einstein-Hilbert action, with the usual notation $\kappa = 8 \pi G c^{-4}$ and $\sqrt{g} \equiv \sqrt{\text{det} g^{IJ}}$, accompanied by the  GW scalar action. The last term $S_{\text{branes}}$ contains interactions localized on the boundaries of the space, consisting of the brane actions in Eq.~\eqref{Act:brane}, as required by the jump conditions at the boundaries $y_i = \{0, L\}$\cite{Csaki:2000zn, Cai:2021mrw}.  By adjusting $V(\phi)$ and $\lambda_i(\phi)$ in this set up,   the bulk scalar  develops a $y$-dependent vacuum expectation value (VEV), $\langle \phi \rangle = \phi_0(y)$, which reacts  on the metric such that the radion field acquires its mass\cite{Goldberger:1999uk, Goldberger:1999un, Csaki:2000zn, Kofman:2004tk}.  

We can now demonstrate the invariance of Eq.~\eqref{Act} under 5d diffeomorphism.  A diffeomorphism involves a pushforward followed by  coordinate transformation back.  Under the  pushforward  $X^I \to X^I + \xi^I(X)$,   an infinitesimal  variation of tensor fields  is generated by Lie derivative.   Since the action in Eq.~\eqref{Act} merely depends on the metric $g_{MN}$ and the GW scalar $\phi$, we will start with the corresponding transformations:
\begin{eqnarray} 
&& \delta g^{MN} = - \left(\nabla^M \xi^N + \nabla^N \xi^M \right)\,,  \label{dg5} \\
&&  \delta \phi = \xi^K \partial_K \phi  \,, \label{dphi} 
\end{eqnarray}
where $\nabla^M \equiv g^{MN} \nabla_N$ denotes the covariant derivative and $\xi^M$ is an arbitrary function in the bulk that vanishes at the boundaries $y=0$ and $y=L$. Assuming $S_{\rm bulk} = \int d^5 x \sqrt{g} \, \mathcal{ L} $,  with   $\mathcal{ L}$ being a scalar constructed out of tensors (as is the case for Eq.~\eqref{Act}),  the  transformations in Eqs.~(\ref{dg5}-\ref{dphi}) lead to the main result: 
\begin{equation} 
\delta \left( \sqrt{g} \mathcal{ L} \right) = \partial_M \left( \xi^M \sqrt{g} \mathcal{ L} \right)\,. \label{variation}
\end{equation}
The proof of the above result requires  two steps of work. 
Firstly, it is easy to prove that the square root of the metric determinant $\sqrt{g}$  transforms as a total derivative under the infinitesimal diffeomorphism, i.e.
\begin{eqnarray}
\delta  \sqrt{g} &=& -\frac{1}{2} \sqrt{g} g_{MN} \left(  \xi ^K  \partial_K g^{M N}-  g^{N K} \partial_{K} \xi^M - g^{MK} \partial_K\xi^N\right) \nonumber \\
&=&    \xi^M \partial_M \sqrt{g}  +\sqrt{g} \partial_M \xi^M 
= \partial_M \left( \xi^M \sqrt{g} \right)\,,   \label{dg} 
\end{eqnarray}
where  we  used the identity $\partial_M \sqrt{g} = - \frac{1}{2}   \sqrt{g}   g_{IJ} \partial_M g^{IJ}$.  
Secondly,  we need  to prove that  the  transformation  of   Lagrangian density  is  a directional derivative, i.e. $\delta \mathcal{ L} = \xi^M \partial_M \mathcal{L}$. In the following we explicitly show that  each term in Eq.~\eqref{Act}  observes this property.
\begin{itemize}

\item[(1)]
We will start with the variation of  Ricci scalar:
\begin{eqnarray}
\delta \mathcal{R}  = \delta g^{MN}  \mathcal{R}_{MN} +  g^{MN} \delta \mathcal{R}_{MN}\,.  \label{eq:R0} 
\end{eqnarray}
Our strategy is to  recast  Eq.~\eqref{eq:R0} in terms of  covariant derivatives.  Using Eq.~\eqref{dg5}, the first term can be expressed as:
\begin{eqnarray}
\delta g^{MN} \mathcal{R}_{MN} =  - \left(\nabla^M \xi^N + \nabla^N \xi^M \right) \mathcal{R}_{MN}\,. \label{eq:R1}
\end{eqnarray}
After a lengthy algebraic manipulation,  the second term in Eq.~\eqref{eq:R0} can be recasted into:
\begin{eqnarray}
g^{MN} \delta \mathcal{R}_{MN} &=& \nabla_M \nabla_N \left( - \delta g^{MN} + g^{MN} g_{IJ} \, \delta g^{IJ} \right) \nonumber \\
&=&  \nabla_M \nabla_N \left(\nabla^M \xi^N + \nabla^N \xi^M - 2 g^{MN} \nabla^K \xi_K \right) \nonumber \\
&=& 2 \nabla^M \left( R_{M N} \xi^N \right)\,,  \label{eq:R2}
\end{eqnarray}
where $ \left(\nabla_M \nabla_N - \nabla_N \nabla_M \right) \xi^M = \mathcal{R}_{M N} \xi^M $ is applied to the last step.
Substituting  Eqs.~(\ref{eq:R1}-\ref{eq:R2}) into Eq.~\eqref{eq:R0},   we find that:
\begin{eqnarray}
\delta \mathcal{R}  = 2 \, \xi^M \nabla^N \mathcal{R}_{MN}\,. 
\end{eqnarray}
Finally, we use the contracted Bianchi identity $\nabla^N \mathcal{R}_{MN} = \frac{1}{2} \nabla_M \mathcal{R} $  to simplify the result:
\begin{eqnarray}
\delta \mathcal{R}  =  \xi^M \nabla_M  \mathcal{R} = \xi^M \partial_M \mathcal{R}\,. \label{dR}
\end{eqnarray}

\item[(2)]  Using Eqs.~(\ref{dg5}-\ref{dphi}), the variation of  the scalar kinetic term can be derived  straightforwardly:
\begin{eqnarray}
\delta \left(  g^{MN} \partial_M  \phi  \partial_N  \phi  \right) &=&  2 g^{MN} \partial_M \phi  \, \partial_N \left(\xi^K \partial_K \phi \right) + \xi^K \partial_K g^{MN} \partial_M \phi \partial_N \phi 
- 2 g^{M K} \partial_K \xi^N \partial_M \phi \partial_N \phi \nonumber \\
&=& \xi^K \partial_K \left( g^{MN} \partial_M \phi \partial_N \phi  \right)\,. \label{dkin}
\end{eqnarray}

\item[(3)]  As  $\phi$ is a fundamental scalar,  the variation of the GW potential  $V(\phi)$ is simply  a directional derivative:
\begin{eqnarray}
\delta V(\phi) = \frac{\partial V}{\partial \phi}  \delta \phi = \xi^K  \partial_K V(\phi)\,. \label{dV}
\end{eqnarray}

\end{itemize}

Combining Eqs.(\ref{dR}-\ref{dV}) with Eq.(\ref{dg}),  one can deduce that $\sqrt g \mathcal{ L}$  indeed  transforms as a total derivative under the infinitesimal diffeomorphism. Therefore the variation of 5d action  becomes a surface term
\beq
\delta S_{\rm bulk} \equiv \delta\left(\int d^5 x \sqrt g  \mathcal{ L} \right)  =  \int d^5 x \partial_M \left( \xi^M \sqrt{g} \mathcal{ L}  \right) =0 \, \label{invariance}
\eeq 
that vanishes as $\xi^M =0$ at the boundaries of the 5d space.

\vspace{0.3cm}
Having demonstrated the invariance of the action  under a general diffeomorphism, we  focus our attention on the  dynamical perturbations around the vacuum of the theory.
When  only the VEVs of the metric and of the GW scalar are taken into account in Eq.(\ref{dg5}-\ref{dphi}), one obtains the linearized diffeomorphism transformations.  However, the linear approximation is not valid when one expands the action beyond the quadratic order.  This is due to the fact that the exact diffeomorphism transformation of the fields contains  a  nonlinear part, which is crucial in rendering the full invariance of the action. Here we will derive the exact transformation of the metric perturbation fields from  the  covariant diffeomorphism  in Eq.~(\ref{dg5}).  
Without loss of generality, we parametrize the line element in the RS model as\cite{Charmousis:1999rg,Csaki:2000zn}:
\beq
d s^2  =  e^{-2A -2F}  \hat g_{\mu \nu} d x^\mu d x^\nu - \left[ 1+G \right]^2 dy^2\,,  \label{metric} \;\; \mbox{where}\;\; \hat g_{\mu \nu} = \eta_{\mu \nu} + h_{\mu \nu}\,. \label{eq:ds2}
\eeq
The  metric  observes  a $S^1/\mathbb{Z}_2$ orbifold symmetry and $\eta_{\mu \nu} = \left(+, -,-,-\right)$ is for the 4d Minkowski spacetime. The graviton degrees of freedom are contained in $h_{\mu\nu} (x,y)$, while the radion is described by the functions $F(x,y),\ G(x,y)$, with $A(y)$ being  the warp factor. Note that  this is most general parametrization in the unitary gauge, which decouples the graviton from the radion.

Due to  the fact  that $g^{MK} g_{NK} = \delta^M_N$, Eq.(\ref{dg5}) can be rewritten as:
\beq 
\delta g_{MN}  = \nabla_M   \xi_N +  \nabla_N   \xi_M  \label{dg1}
\eeq
with $\xi_M = g_{MN} \xi^N$. For convenience, we  split the gauge parameter into two parts: $\xi^\mu = \hat \xi^\mu $ and $\xi^5 = \epsilon$.  As we demonstrated above, as long as the variation of the metric observes Eq.~(\ref{dg1}), the Einstein-Hilbert action is invariant under the diffeomorphism transformation. Hence, in order to derive the exact transformation rules for  the perturbation fields  in the unitary gauge, i.e. $g_{\mu 5} =0$, we will not adopt any approximation. Evaluating the right hand side of Eq.~(\ref{dg1}),  its 4d  components read:
\beq
\nabla_\mu   \xi_\nu  =   \partial_\mu \left(g_{\nu \rho} \hat \xi^\rho \right) - \Gamma^{N}_{\mu \nu}  g_{N M}  \xi^M 
= g_{\nu \rho} \partial_\mu  \hat \xi^\rho + \frac{1}{2} \left[  \partial_\mu g_{\rho \nu} - \partial_\nu g_{\rho \mu}   \right]  \hat \xi^\rho +  \frac{1}{2} \ \xi^M \partial_M g_{\mu \nu} \,,       \label{cd1}
\eeq
where we recall that the sums over $M, N$ span over all the dimensions, while  the Greek indices are confined to the 4d Minkowski spacetime. Note that, although  $g_{\mu 5} =0$ is imposed on Eq.~(\ref{cd1}),  the last term still includes a dependence on  $\xi^5 = \epsilon$,  that originates  from the Christoffel connection. Similarly, the fifth component of Eq.~(\ref{dg1})  reads:
\beq
\nabla_5   \xi_5  =   \partial_5 \left(g_{55}  \epsilon \right) - \Gamma^{N}_{55}  g_{N M}  \xi^M =  g_{55}  \partial_5 \epsilon  +  \frac{1}{2}  \xi^M \partial_M g_{55}\,.  \label{cd2}
\eeq
Substituting Eqs.~(\ref{cd1}-\ref{cd2}) into Eq.~{\ref{dg1}}, one obtains:
\beq
\delta g_{\mu \nu} &=&  g_{\mu \rho} \partial_\nu  \hat \xi^\rho + g_{\nu \rho} \partial_\mu  \hat \xi^\rho + \xi^M \partial_M g_{\mu \nu}\,,  \nonumber  \label{dg4a} \\
\delta g_{55} &=&  2 \, g_{55} \, \partial_5 \epsilon   + \xi^M \partial_M g_{55}\,.  \label{dg5a}
 \eeq
Finally, using the explicit form of the metric, the variations of the metric on the left hand side of Eq.(\ref{dg1}) yields:
 \beq
\delta g_{\mu \nu}  &=& e^{-2A- 2F} \delta h_{\mu \nu} - 2 \, e^{-2A -2F} \delta F \hat g_{\mu \nu} \,, \nonumber \label{dg4b} \\
\delta g_{55} &=&  -2 (1+G) \, \delta G \,. \label{dg5b}
\eeq
By comparing Eqs.(\ref{dg5a}) and Eqs.(\ref{dg5b}), we can extract  the transformation rules for the fields:
\beq
\delta h_{\mu \nu} &=&   \left(\partial_{\mu} \hat{\xi}_{\nu} + \partial_\nu \hat{\xi}_\mu  \right)  
+ \hat{\xi}^\alpha \partial_\alpha  h_{\mu \nu} +  \partial_\mu \hat{\xi}^\alpha h_{\alpha \nu} + \partial_\nu \hat{\xi}^\alpha h_{\alpha \mu} + \epsilon   h'_{\mu \nu}\,,  \label{hrule}  \\
\delta F &=&  A'  \epsilon  +\epsilon F' +\hat{\xi}^\alpha \partial_\alpha F \,, \, \label{Frule} \\ 
\delta G  &=&  \epsilon' +\partial_5 \left( \epsilon G \right) +\hat{\xi}^\alpha \partial_\alpha G\,,  \,  \label{Grule}
\eeq
with $\hat \xi_\mu = \eta_{\mu \nu} \hat \xi^\nu$ and the prime standing for $\partial_5 = \partial/\partial y$. In fact, the linearized result\cite{Kogan:2001qx} is related to Eq.(\ref{hrule}-\ref{Grule}) by $\epsilon =W'(x,y) e^{-2A}$, where  $\xi^M = (\hat \xi^\mu, \epsilon)$  are functions of  $(x^\mu,y)$ to make the linear variation of $g_{\mu 5}$ to vanish. But that setup does not work at the nonlinear level (see Eq.(\ref{du5})).  In Appendix~\ref{Appendix}, we also derive the exact diffeomorphism transformation in the conformal coordinate, which confirms that the choice of coordinate does not matter.   Similarly, in terms of  $\hat \xi^\mu$ and $\epsilon$,  the transformation for the GW scalar  reads:
\beq
\delta \varphi = \epsilon  \phi'_0  + \epsilon  \varphi' + \hat \xi^\alpha \partial_\alpha \varphi  \label{Vrule} \,.
\eeq
Eqs.~(\ref{hrule}-\ref{Vrule}) indicate  that the diffeomorphism acts as a re-parametrization symmetry that does not mix physical fields with different spins. Furthermore, the gauge parameters $\hat{\xi}^\mu$ and $\epsilon$ must respect certain constraints from the unitary gauge, as  the transformation should  keep the metric in its  original form.  Because $g_{55}$ and $g_{\mu \nu} $ are field dependent, this  results in:
\beq
 \delta g_{\mu 5} & \, = \, g_{55} \partial_\mu \epsilon + g_{\mu \nu}  \partial_5 \hat \xi^\nu =0 \;\;
 \Rightarrow \;\;  \partial_\mu \epsilon =0   \quad  \mbox{and}  \quad  \partial_5 \hat \xi^\nu =0\,, \label{du5}
\eeq
which implies  that $\epsilon$ is a  function of a single variable $y$ while $\hat \xi^\mu$  depends only on the Minkowski coordinates.  Note that  the diffeomorphism transformations in Eqs.~(\ref{hrule}-\ref{Grule}) are observed by off-shell fields: the on-shell conditions  $\partial^\mu h_{\mu \nu} = h =0$ or $G= 2F$, in fact, are not preserved due to the nonlinearity.

The main result of our work is that the off-shell symmetry defines a recursive  relation  among consecutive terms in the Lagrangian expansion. Let's first expand the  Lagrangian density  in the bulk action  $S_{\rm bulk} = \int d^5 x \sqrt{g} \, \mathcal{ L} $ as
\beq
\sqrt{g} \  \mathcal{ L} = \ \sum_n  \mathcal{ \hat L}^{(n)} \,,
\eeq
with  $\mathcal{\hat L}^{(n)} $ being the $n$-th order Lagrangian expansion in powers of the fields $h^{\mu\nu}$, $F$, $G$ and $\varphi$. Note that it also contains the metric perturbation contribution from $\sqrt{g}$.  Due to  the orbifold symmetry,  the $A''$ or $\phi''_0$ term in $\mathcal{\hat L}^{(n)}$  generates  boundary  contributions proportional to $\delta(y-y_i)$. However as we show in Appendix ~\ref{app:n=1}, these boundary terms need not be subtracted out as long as $\epsilon(y_i) =0$ is imposed when the fifth dimensional diffeomorphism is invoked.
Schematically, the variation of the fields under diffeomorphism can be split as 
\beq
\delta =  \left. \delta^{(1)} \right|_{\rm linear}  + \left. \delta^{(2)}\right|_{\rm nonlin}\,,
\eeq
where $\delta^{(1)}$ contains the linear terms independent on the perturbation fields (i.e. the first terms in the left hand side of Eqs.~(\ref{hrule}-\ref{Vrule}), while $\delta^{(2)}$ contains the nonlinear terms with $\epsilon$ or $\hat \xi^\mu$ multiplying the fields (all the remaining terms in Eqs.~(\ref{hrule}-\ref{Vrule}). Hence, one can easily show that the relation  $\delta \left( \sqrt{g} \mathcal{ L}  \right)= \partial_M \left( \xi^M \sqrt{g} \mathcal{ L} \right)$  holds true  if and only if  the  Lagrangian expansion terms  observe the following recursive relation:
\beq  
\delta^{(2)} \mathcal{\hat L}^{(n)} +  \delta^{(1)} \mathcal{\hat L}^{(n+1)}  =  \partial_M \left( \xi^M  \mathcal{\hat L}^{(n)} \right)\,. \label{recursive}
\eeq
which states that  summing  the linear transformation of the $(n+1)$-th order term with the nonlinear transformation of the $n$-th order term,  yields  a total derivative containing the $n$-th (lower) order term.  Note that Eq.(\ref{recursive}) is valid without performing the  5d integration, hence  any non-dynamical surface term arising from the Lagrangian expansion must be retained.  In the absence of  ambiguity associated with 5d integration,  the recursive relations impose nontrivial constraints on the interaction structure of gravitons  in the RS model.  

\vspace{0.3cm}
We have shown so far that the bulk action in Eq.~\eqref{Act} for the RS model with GW stabilization is invariant under an off-shell symmetry. Note that this also requires $F, G, \varphi$ transform independently. The most striking consequence of this exact diffeomorphism invariance  is encoded in the recursive relations Eq.~\eqref{recursive}.
To better elucidate how these relations act in practice, we first focus on the simplest $n=0$ case.
Expanding the bulk Lagrangian up to linear order in the fields~\cite{Cai:2021mrw}, and keeping all the total derivatives, the $n=0$ field-independent term reads
\beq
\mathcal{\hat L}^{(0)}  = -\frac{4}{\kappa^2} e^{-4A} \left(  A^{\prime 2} -  A'' \right) -  e^{-4A} \phi'^2_0\,.
\eeq
where the $\phi'^2_0$ term includes the contribution from $-e^{-4A}V(\phi_0)= -e^{-4A} \left(  \frac{1}{2} \phi^{\prime 2}_0 -\frac{6}{\kappa^2} A'^2 \right)$ and $\frac{1}{2} e^{-4A}\eta^{55}(\partial_5 \phi_0)^2$. while the $n=1$ term, linear in the fields, can be written in terms of  two pieces:
\beq
\mathcal{\hat L}_{kin}^{(1)}  &=&  -\frac{1}{2 \kappa^2}   e^{-2A} (\partial_{\mu} \partial_\nu h^{\mu \nu}  -\Box h )  -\frac{1}{\kappa^2}   e^{-2A} \Box (3F-G) \,, \label{eq:Lkin(1)}\\
\mathcal{\hat L}_{pot}^{(1)}  &=&-\frac{1}{2 \kappa^2}   \left[   \partial_5 \left[ e^{-4A} \left( h' -  A' h \right) \right] -  e^{-4 A}  h \left[ 3  A'' - \kappa^2   \phi^{\prime 2}_0 \right] \right] +\nonumber  \\ 
&& \frac{4}{\kappa^2}    \partial_5 \left[ e^{-4A}  (F'- A'G- A' F)\right] -  \partial_5\left[ e^{-4A} \phi'_0 \varphi \right] - \nonumber \\ && e^{-4A} \sum_i \left(4F \lambda(\phi_0)- \frac{\partial \lambda_i}{\partial \phi_0} \varphi \right) \delta (y- y_0)\,, \label{eq:Lpot(1)}
\eeq
where $kin$ and $pot$ label the terms containing four dimensional derivatives or not,  respectively.  Eq.(\ref{eq:Lkin(1)}-\ref{eq:Lpot(1)}) include both the graviton and radion contributions, with their derivation details given in Appendix~\ref{app:tad}. As $\mathcal{\hat L}^{(0)}$ does not contain 4d derivatives nor fields, the recursive relation implies that
\beq
\delta^{(1)} \mathcal{\hat L}_{pot}^{(1)} = \partial_M \left(\xi^M \mathcal{\hat L}^{(0)}\right)\,, \label{eq:dL1}
\eeq
indicating that the linear variation of the linear potential term by itself is a total derivative. As a consequence,
\beq
\delta^{(1)} \mathcal{\hat L}_{kin}^{(1)} = 0\,. \label{eq:dL2}
\eeq
To check  Eqs.~(\ref{eq:dL1}-\ref{eq:dL2}), one can split the linear variation $\delta^{(1)} = \delta^{(1)}_{\epsilon} + \delta^{(1)}_{\xi} $ with respect to gauge parameters $\xi^\mu$ and $\epsilon$.  Starting from the kinetic term in Eq.~\eqref{eq:Lkin(1)} and using the linear part of Eq.~\eqref{hrule}, we obtain:
\beq
\delta_{\xi}^{(1)} \mathcal{\hat L}^{(1)}_{kin} &=&-\frac{1}{2\kappa^2} e^{-2A} \delta^{(1)} \left(  \partial^\mu \partial^\nu h_{\mu \nu } -\Box h \right)  \nonumber \\ &=& -\frac{1}{2\kappa^2} e^{-2A}\left[ \partial^\mu \partial^\nu (\partial_\mu \xi_\nu + \partial_\nu \xi_\mu ) - 2 \Box \partial_\mu \xi^\mu \right] =0\,.
\eeq
Similarly, for the variation under $\epsilon$, using the linear part of Eqs.~(\ref{Frule}--\ref{Grule}), we obtain:
\beq
\delta_{\epsilon}^{(1)} \mathcal{\hat L}^{(1)}_{kin} = -\frac{1}{\kappa^2} e^{-2A} \delta^{(1)}\left[ \Box \left( 3F - G\right)  \right] = -\frac{1}{\kappa^2} e^{-2A}\Box \left( 3 A' \epsilon - \epsilon' \right) = 0\,.
\eeq
For the potential term, we can again decompose the linear variation  to find:  
\beq
\delta_{\epsilon}^{(1)} \mathcal{\hat L}_{pot}^{(1)} &=& -\frac{4}{ \kappa^2} \partial_5  \left[e^{-4A} \left(\partial_5 (\epsilon A') - A' \epsilon' - A'^2 \epsilon \right)\right]  -  \partial_5\left[ e^{-4A} \epsilon \phi^{\prime 2}_0\right] = \partial_5 \left( \epsilon  \mathcal{\hat L}^{(0)}  \right) 
\eeq
\beq
\delta_{\xi}^{(1)} \mathcal{\hat L}_{pot}^{(1)} &=& -\frac{1}{ 2\kappa^2}   \left[   \partial_5 \left[ - e^{-4A} \left(  2A' \partial_\mu \hat \xi^\mu   \right) \right] - 2  e^{-4 A}  \partial_\mu \hat \xi^\mu \left[ 3  A'' - \kappa^2   \phi^{\prime 2}_0 \right] \right]   =  \partial_\mu \left( \hat \xi^\mu  \mathcal{\hat L}^{(0)}  \right) \label{eq:dL3}
\eeq
The property above stems from the fact that the linear transformation of $h^{\mu\nu}$ only contains $\hat \xi^\mu$, while that for $F$, $G$ and $\varphi$ only contains $\epsilon$. Note that the boundary term proportional to $\delta(y-y_i)$in the last line of Eq.(\ref{eq:Lpot(1)}) does not contribute due to $\epsilon(y_i) =0$ at the fixed points. In fact,  the term in the form of $\left[ 3  A'' - \kappa^2   \phi^{\prime 2}_0 \right] =  \kappa^2\sum_i  \lambda_i (\phi_0) \delta(y-y_0)  $ in Eq.(\ref{eq:dL3}) is also a boundary term, but its contribution should be included as the result of  4d diffeomorphism.

\vspace{0.3cm}
To highlight the significance of nonlinear variations emerging from the $n=1$ recursive relation,  we  focus on the explicit proof involving kinetic terms. Let us  write down the second order bulk Lagrangian with two $\partial_\mu$ derivatives, following the derivation in \cite{Cai:2022geu}:
\beq
\mathcal{\hat L}_{kin}^{(2)} &=&  -\frac{1}{ 2\kappa^2}  e^{-2A} \left[ {\mathcal L}_{FP} + \partial_\alpha \left(  h_{\mu \nu} \partial^\alpha h^{\mu \nu } +  \partial_\beta \left(h^{\alpha \beta} h- h^{\alpha \nu}  h^\beta_\nu \right)  - \frac{1}{2} h \partial_\beta h^{\alpha \beta} -\frac{1}{2} h \partial^\alpha h \right)\right] 
\nonumber \\
&& -\frac{1}{ \kappa^2}  e^{-2A} \Big[  3\partial_\mu F \partial^\mu \left( F - G\right) +  \partial_\mu \Big(  3( G-2F)  \partial^\mu F + 2 F \partial^\mu G   \Big)\Big] + \mathcal{L}_{kin, mix}^{(2)}  
\eeq 
where the Fierz-Pauli Lagrangian reads ${\mathcal L}_{FP} = \frac{1}{2} \partial_\nu h_{\mu \alpha} \, \partial^\alpha    
h^{\mu \nu} -   \frac{1}{4} \partial_\mu h_{\alpha \beta} \, \partial^\mu h^{\alpha \beta}    
- \frac{1}{2} \partial_\alpha h \, \partial_\beta h^{\alpha \beta} +\frac{1}{4}    
\partial_\alpha h \, \partial^\alpha h$. We amend the conventional kinetic terms for graviton and radion with non-dynamical total derivative terms,  which is crucial to prove the the recursive relations in Eq.~\eqref{recursive}.  And there exists a mixing quadratic  with terms linear in both $h$ and the radion fields:
\beq
\mathcal{L}_{kin, mix}^{(2)}  &=& -\frac{1}{2 \kappa^2} e^{-2A} \left( G- 2F\right) \left[ \partial_\mu \partial_\nu h^{\mu \nu} -\Box h \right] -\frac{1}{\kappa^2} e^{-2A} \partial_\mu \left[ \left(\frac{1}{2}  h \eta^{\mu\nu} - h^{\mu \nu}\right)  \partial_\nu (3F-G) \right]
\eeq
Since each component in $\mathcal{L}_{kin}^{(1)}= \mathcal{L}^{(1)}_{kin, rad} + \mathcal{L}^{(1)}_{kin, h}$ can be nonlinearly varied under $\epsilon$ or $\hat \xi^\mu$, the $n=1$ recursive relation actually comprises four independent parts. Firstly, we consider the relevant variations of radion kinetic terms  under the fifth dimensional diffeomorphism proportional to $\epsilon$, which  gives:
\beq 
&&  \delta_{\epsilon}^{(2)}  \mathcal{L}^{(1)}_{kin, rad} + \delta_{\epsilon}^{(1)}  \mathcal{L}^{(2)}_{kin, rad}  \nonumber \\ &=&  -\frac{1}{\kappa^2 } e^{-2A} \Big[ \left[ \epsilon \partial_5 \Box \left( 3F-G \right)  - \epsilon' \Box G \right]  + \left[ 3 \epsilon' \Box F + 2 A' \epsilon \Box \left( G - 3F \right) \right]  \Big]  \nonumber \\ 
& =&  -\frac{1}{\kappa^2 } \partial_5 \left[ \epsilon \, e^{-2A}  \Box\left( 3F -G\right) \right] = \partial_5 \left[ \epsilon  \,  \mathcal{L}^{(1)}_{kin, rad} \right]\,,    \label{eq:dkin1}
\eeq
that  is equal to the total  fifth dimensional derivative of the linear radion  term.  And in the linear variation of $\mathcal{L}^{(2)}_{kin, rad} $, only terms with the radion fields not operated with 4d derivative could survive. Note that $\delta_\xi^{(2)}  \mathcal{L}^{(1)}_{kin, rad} $  shares the same structure as $  \delta^{(1)}_\xi \mathcal{L}_{kin, mix}^{(2)} $, the summation of these two terms leads to:
\beq
&&  \delta_\xi^{(2)}  \mathcal{L}^{(1)}_{kin, rad} +  \delta^{(1)}_\xi \mathcal{L}_{kin, mix}^{(2)}  \nonumber \\ &=&-\frac{1}{\kappa^2 } e^{-2A} \Big[\Box \left[\hat \xi^\mu \partial_\mu  (3 F-G)  \right] + \partial_\mu \left[ \left(\partial_\alpha  \hat \xi^\alpha \eta^{\mu\nu} -\partial^\mu \hat \xi^\nu- \partial^\nu \hat \xi^\mu\right)  \partial_\nu (3F-G) \right] \Big] \nonumber \\
&=& -\frac{1}{\kappa^2 } e^{-2A}  \partial_\mu \left[\hat \xi^\mu  \Box (3 F-G)  \right] = \partial_\mu \left[ \hat\xi^\mu  \,  \mathcal{L}^{(1)}_{kin, rad} \right] \,, \label{eq:dkin2}
\eeq
Then we should consider the  variation involving the graviton kinetic terms.  With tedious but straightforward calculation, we  can verify that:
\beq
\delta^{(2)}_{\xi} \mathcal{L}_{kin, h}^{(1)} + \delta^{(1)}_{\xi}  \mathcal{L}_{kin, h^2}^{(2)} = \partial_\mu \left[\hat \xi^\mu  \mathcal{L}_{kin, h}^{(1)}\right]  \,. \label{eq:dkin3}
\eeq
In fact, this result  is  valid in the 4d  gravity theory without a radion field.  Finally we go through  the variation of  graviton terms with respect to $\epsilon$:
\beq
 \delta_\epsilon^{(2)}  \mathcal{L}^{(1)}_{kin, h}  +  \delta^{(1)}_\epsilon \mathcal{L}_{kin, mix}^{(2)} &=&  -\frac{1}{2\kappa^2} e^{-2A} \left[ \epsilon(\partial_{\mu} \partial_\nu  h^{\prime \mu \nu}  -\Box h' )+ \left( \epsilon' -2 A' \epsilon \right) (\partial_{\mu} \partial_\nu h^{\mu \nu}  -\Box h )\right] 
 \nonumber \\ & =&   -\frac{1}{2\kappa^2}  \partial_5 \Big[ \epsilon  e^{-2A} \left(\partial_{\mu} \partial_\nu h^{\mu \nu}  -\Box h \right) \Big]   = \partial_5 \left[ \epsilon  \,  \mathcal{L}^{(1)}_{kin, h} \right] \,.  \label{eq:dkin4}
\eeq
Combining Eqs.(\ref{eq:dkin1}-\ref{eq:dkin4}), we proved $\delta^{(2)} \mathcal{L}_{kin}^{(1)} + \delta^{(1)}  \mathcal{L}_{kin}^{(2)} = \partial_M \left(\xi^M  \mathcal{L}_{kin}^{(1)}\right)$ as predicted from Eq.~\eqref{recursive}. One can  refer to Appendix~\ref{app:n=1} for the $n=1$ recursive relations of  potential terms.  Higher order recursive relations impose non-trivial constraints on the interaction structure. The case $n=2$ was explicitly verified in Ref.\cite{Cai:2023mqn} via the trilinear interaction involving gravitons and radions.

\vspace{0.3cm}
Now we comment on  the  property  of this symmetry. When we discuss the off-shell diffeomorphism,  all the scalar fields $F, G, \varphi$ are assumed to be independent.  While prior to GW stabilization,  the RS action with conformal metric ($A''=0$ in the bulk) is invariant under an on-shell diffeormorphism. The on-shell condition is a linearized Einstein equation: $F'-A' G =0$. Examining the action of the diffeomorphism on $F$ and $G$  using Eqs.(\ref{Frule}-\ref{Grule}), we find that:  
\beq
 \delta \left( F'  - A'  G \right) =   \partial_5 \left[ \epsilon (F' - A' G )\right]  + \hat \xi^\alpha \partial_\alpha \left( F' - A' G  \right) +
 A''\epsilon (1+G)\,. \label{constraint}
\eeq
Imposing the EOM, $F'-A'G =0$, and the condition $\epsilon(y_i) =0 $  on the boundaries, the above equation reduces to $\delta \left( F' - A'  G\right) = 0$ for $A''=\frac{\kappa^2}{3} \sum_i  \lambda_i (\phi_0) \delta(y-y_0) $. Thus, substituting $G = F'/A'$  into the 5d action effectively removes one degree of freedom, while keeping the constrained action  invariant under diffeomorphism. However, the GW mechanism breaks this on-shell diffeomorphism and gives mass to the radion, as the corresponding EOM  is modified to  be $F'  - A'  G = \frac{\kappa^2}{3} \phi'_0 \varphi$ for  a massive radion.
This implies that if we eliminate $G$ from the 5d action with the $\varphi$-dependent EOM,  the invariance is no long valid.  In fact,  this property can  be considered as the consequence of  radion stabilization. Also the VEV $\phi_0$ is  actually  an order parameter in the confinement phase transition~\cite{Creminelli:2001th,Nardini:2007me,Konstandin:2010cd,Agashe:2020lfz},  and  the radion dynamics plays a crucial role in determining the spectrum of stochastic gravitational waves. 

\vspace{0.3cm}
To summarize, we have proved that the RS  bulk action remains invariant under diffeomorphism even after the Goldberger-Wise stabilization field is included, provided the transformation operates on off-shell fields. In particular, we show that the fifth-dimensional diffeomorphism is consistent with conformal symmetry in a pure AdS slice, if the scalar fields $F$ and $G$ need to satisfy an appropriate relation.  A key consequence of this invariance is the emergence of a set of recursive relations linking consecutive orders of Lagrangian expansion.  These relations provide a systematic tool for determining the  interaction structure beyond the quadratic order in the RS model.

\bibliographystyle{JHEP}
\bibliography{warped}

\providecommand{\href}[2]{#2}\begingroup\raggedright\begin{thebibliography}{10}

\bibitem{Randall:1999vf}
L.~Randall and R.~Sundrum, \emph{{An Alternative to compactification}},
  \href{https://doi.org/10.1103/PhysRevLett.83.4690}{\emph{Phys. Rev. Lett.}
  {\bfseries 83} (1999) 4690}
  [\href{https://arxiv.org/abs/hep-th/9906064}{{\ttfamily hep-th/9906064}}].

\bibitem{Randall:1999ee}
L.~Randall and R.~Sundrum, \emph{{A Large mass hierarchy from a small extra
  dimension}}, \href{https://doi.org/10.1103/PhysRevLett.83.3370}{\emph{Phys.
  Rev. Lett.} {\bfseries 83} (1999) 3370}
  [\href{https://arxiv.org/abs/hep-ph/9905221}{{\ttfamily hep-ph/9905221}}].

\bibitem{Maldacena:1997re}
J.M.~Maldacena, \emph{{The Large N limit of superconformal field theories and
  supergravity}}, \href{https://doi.org/10.1023/A:1026654312961}{\emph{Adv.
  Theor. Math. Phys.} {\bfseries 2} (1998) 231}
  [\href{https://arxiv.org/abs/hep-th/9711200}{{\ttfamily hep-th/9711200}}].

\bibitem{Witten:1998qj}
E.~Witten, \emph{{Anti-de Sitter space and holography}},
  \href{https://doi.org/10.4310/ATMP.1998.v2.n2.a2}{\emph{Adv. Theor. Math.
  Phys.} {\bfseries 2} (1998) 253}
  [\href{https://arxiv.org/abs/hep-th/9802150}{{\ttfamily hep-th/9802150}}].

\bibitem{Gubser:1998bc}
S.S.~Gubser, I.R.~Klebanov and A.M.~Polyakov, \emph{{Gauge theory correlators
  from noncritical string theory}},
  \href{https://doi.org/10.1016/S0370-2693(98)00377-3}{\emph{Phys. Lett. B}
  {\bfseries 428} (1998) 105}
  [\href{https://arxiv.org/abs/hep-th/9802109}{{\ttfamily hep-th/9802109}}].

\bibitem{Batell:2008me}
B.~Batell, T.~Gherghetta and D.~Sword, \emph{{The Soft-Wall Standard Model}},
  \href{https://doi.org/10.1103/PhysRevD.78.116011}{\emph{Phys. Rev. D}
  {\bfseries 78} (2008) 116011}
  [\href{https://arxiv.org/abs/0808.3977}{{\ttfamily 0808.3977}}].

\bibitem{Cabrer:2009we}
J.A.~Cabrer, G.~von Gersdorff and M.~Quiros, \emph{{Soft-Wall Stabilization}},
  \href{https://doi.org/10.1088/1367-2630/12/7/075012}{\emph{New J. Phys.}
  {\bfseries 12} (2010) 075012}
  [\href{https://arxiv.org/abs/0907.5361}{{\ttfamily 0907.5361}}].

\bibitem{Gherghetta:2010he}
T.~Gherghetta and N.~Setzer, \emph{{On the stability of a soft-wall model}},
  \href{https://doi.org/10.1103/PhysRevD.82.075009}{\emph{Phys. Rev. D}
  {\bfseries 82} (2010) 075009}
  [\href{https://arxiv.org/abs/1008.1632}{{\ttfamily 1008.1632}}].

\bibitem{Randall:2006py}
L.~Randall and G.~Servant, \emph{{Gravitational waves from warped spacetime}},
  \href{https://doi.org/10.1088/1126-6708/2007/05/054}{\emph{JHEP} {\bfseries
  05} (2007) 054} [\href{https://arxiv.org/abs/hep-ph/0607158}{{\ttfamily
  hep-ph/0607158}}].

\bibitem{Konstandin:2011dr}
T.~Konstandin and G.~Servant, \emph{{Cosmological Consequences of Nearly
  Conformal Dynamics at the TeV scale}},
  \href{https://doi.org/10.1088/1475-7516/2011/12/009}{\emph{JCAP} {\bfseries
  12} (2011) 009} [\href{https://arxiv.org/abs/1104.4791}{{\ttfamily
  1104.4791}}].

\bibitem{Baratella:2018pxi}
P.~Baratella, A.~Pomarol and F.~Rompineve, \emph{{The Supercooled Universe}},
  \href{https://doi.org/10.1007/JHEP03(2019)100}{\emph{JHEP} {\bfseries 03}
  (2019) 100} [\href{https://arxiv.org/abs/1812.06996}{{\ttfamily
  1812.06996}}].

\bibitem{Goldberger:1999uk}
W.D.~Goldberger and M.B.~Wise, \emph{{Modulus stabilization with bulk fields}},
  \href{https://doi.org/10.1103/PhysRevLett.83.4922}{\emph{Phys. Rev. Lett.}
  {\bfseries 83} (1999) 4922}
  [\href{https://arxiv.org/abs/hep-ph/9907447}{{\ttfamily hep-ph/9907447}}].

\bibitem{Goldberger:1999un}
W.D.~Goldberger and M.B.~Wise, \emph{{Phenomenology of a stabilized modulus}},
  \href{https://doi.org/10.1016/S0370-2693(00)00099-X}{\emph{Phys. Lett. B}
  {\bfseries 475} (2000) 275}
  [\href{https://arxiv.org/abs/hep-ph/9911457}{{\ttfamily hep-ph/9911457}}].

\bibitem{Hinterbichler:2011tt}
K.~Hinterbichler, \emph{{Theoretical Aspects of Massive Gravity}},
  \href{https://doi.org/10.1103/RevModPhys.84.671}{\emph{Rev. Mod. Phys.}
  {\bfseries 84} (2012) 671} [\href{https://arxiv.org/abs/1105.3735}{{\ttfamily
  1105.3735}}].

\bibitem{Pilo:2000et}
L.~Pilo, R.~Rattazzi and A.~Zaffaroni, \emph{{The Fate of the radion in models
  with metastable graviton}},
  \href{https://doi.org/10.1088/1126-6708/2000/07/056}{\emph{JHEP} {\bfseries
  07} (2000) 056} [\href{https://arxiv.org/abs/hep-th/0004028}{{\ttfamily
  hep-th/0004028}}].

\bibitem{Kogan:2001qx}
I.I.~Kogan, S.~Mouslopoulos, A.~Papazoglou and L.~Pilo, \emph{{Radion in
  multibrane world}},
  \href{https://doi.org/10.1016/S0550-3213(02)00009-3}{\emph{Nucl. Phys. B}
  {\bfseries 625} (2002) 179}
  [\href{https://arxiv.org/abs/hep-th/0105255}{{\ttfamily hep-th/0105255}}].

\bibitem{Chivukula:2022kju}
R.S.~Chivukula, E.H.~Simmons and X.~Wang, \emph{{Supersymmetry and sum rules in
  the Goldberger-Wise model}},
  \href{https://doi.org/10.1103/PhysRevD.106.035026}{\emph{Phys. Rev. D}
  {\bfseries 106} (2022) 035026}
  [\href{https://arxiv.org/abs/2207.02887}{{\ttfamily 2207.02887}}].

\bibitem{Csaki:2000zn}
C.~Csaki, M.L.~Graesser and G.D.~Kribs, \emph{{Radion dynamics and electroweak
  physics}}, \href{https://doi.org/10.1103/PhysRevD.63.065002}{\emph{Phys. Rev.
  D} {\bfseries 63} (2001) 065002}
  [\href{https://arxiv.org/abs/hep-th/0008151}{{\ttfamily hep-th/0008151}}].

\bibitem{Cai:2021mrw}
H.~Cai, \emph{{Radion dynamics in the multibrane Randall-Sundrum model}},
  \href{https://doi.org/10.1103/PhysRevD.105.075009}{\emph{Phys. Rev. D}
  {\bfseries 105} (2022) 075009}
  [\href{https://arxiv.org/abs/2109.09681}{{\ttfamily 2109.09681}}].

\bibitem{Kofman:2004tk}
L.~Kofman, J.~Martin and M.~Peloso, \emph{{Exact identification of the radion
  and its coupling to the observable sector}},
  \href{https://doi.org/10.1103/PhysRevD.70.085015}{\emph{Phys. Rev. D}
  {\bfseries 70} (2004) 085015}
  [\href{https://arxiv.org/abs/hep-ph/0401189}{{\ttfamily hep-ph/0401189}}].

\bibitem{Charmousis:1999rg}
C.~Charmousis, R.~Gregory and V.A.~Rubakov, \emph{{Wave function of the radion
  in a brane world}},
  \href{https://doi.org/10.1103/PhysRevD.62.067505}{\emph{Phys. Rev. D}
  {\bfseries 62} (2000) 067505}
  [\href{https://arxiv.org/abs/hep-th/9912160}{{\ttfamily hep-th/9912160}}].

\bibitem{Cai:2022geu}
H.~Cai, \emph{{Effective Lagrangian and stability analysis in warped space}},
  \href{https://doi.org/10.1007/JHEP09(2022)195}{\emph{JHEP} {\bfseries 09}
  (2022) 195} [\href{https://arxiv.org/abs/2201.04053}{{\ttfamily
  2201.04053}}].

\bibitem{Cai:2023mqn}
H.~Cai, \emph{{Diffeomorphism on-shell breaking from radion stabilization}},
  \href{https://arxiv.org/abs/2309.07904}{{\ttfamily 2309.07904}}.

\bibitem{Creminelli:2001th}
P.~Creminelli, A.~Nicolis and R.~Rattazzi, \emph{{Holography and the
  electroweak phase transition}},
  \href{https://doi.org/10.1088/1126-6708/2002/03/051}{\emph{JHEP} {\bfseries
  03} (2002) 051} [\href{https://arxiv.org/abs/hep-th/0107141}{{\ttfamily
  hep-th/0107141}}].

\bibitem{Nardini:2007me}
G.~Nardini, M.~Quiros and A.~Wulzer, \emph{{A Confining Strong First-Order
  Electroweak Phase Transition}},
  \href{https://doi.org/10.1088/1126-6708/2007/09/077}{\emph{JHEP} {\bfseries
  09} (2007) 077} [\href{https://arxiv.org/abs/0706.3388}{{\ttfamily
  0706.3388}}].

\bibitem{Konstandin:2010cd}
T.~Konstandin, G.~Nardini and M.~Quiros, \emph{{Gravitational Backreaction
  Effects on the Holographic Phase Transition}},
  \href{https://doi.org/10.1103/PhysRevD.82.083513}{\emph{Phys. Rev. D}
  {\bfseries 82} (2010) 083513}
  [\href{https://arxiv.org/abs/1007.1468}{{\ttfamily 1007.1468}}].

\bibitem{Agashe:2020lfz}
K.~Agashe, P.~Du, M.~Ekhterachian, S.~Kumar and R.~Sundrum, \emph{{Phase
  Transitions from the Fifth Dimension}},
  \href{https://doi.org/10.1007/JHEP02(2021)051}{\emph{JHEP} {\bfseries 02}
  (2021) 051} [\href{https://arxiv.org/abs/2010.04083}{{\ttfamily
  2010.04083}}].

\end{thebibliography}\endgroup

\newpage
\appendix  
 
\section{Diffeomorphism in conformal coordinate}~\label{Appendix} 
The metric in the RS model  can be parametrized in the context of  the conformal coordinate:
\beq
d^2 s = e^{-2A(z)} \left[ e^{-2F} \left(\eta_{\mu \nu} + h_{\mu \nu}\right) dx^\mu dx^\nu - (1+ G)^2 d^2 z  \right]
\eeq
where  $e^{-A} dz = dy$ is employed to transform from the metric in Eq.(\ref{metric}). In the conformal coordinate, Eq.(\ref{dg4a}-\ref{dg5a}) are explicitly written as:
\beq
\delta g_{\mu \nu} 
& = & g_{\mu \rho} \partial_\nu  \hat \xi^\rho + g_{\nu \rho} \partial_\mu  \hat \xi^\rho +  \hat \xi^\rho \partial_\rho g_{\mu \nu} \nonumber \\
& + & \zeta \left[ \partial_z e^{-2A -2F} \left( \eta_{\mu \nu} + h_{\mu \nu} \right) +  e^{-2A-2F}   \partial_z  h_{\mu \nu} \right] \,   \label{dg4c}
\eeq    
and
\beq
\delta g_{55} 
 &=&  - 2 (1+G)^2 e^{-2A} \partial_z \zeta - \hat \xi^\mu \partial_\mu \left[ (1+G)^2 e^{-2A }\right] \nonumber
 \\ &-&  \zeta \partial_z \left[ \left(1+G \right)^2 e^{-2A}\right] \, \label{dg5c}
\eeq
From Eq.(\ref{dg4c}-\ref{dg5c}), one can extract  out the component field transformation rules:
\beq
\delta h_{\mu \nu} &=& \left(\partial_{\mu} \hat{\xi}_{\nu} + \partial_\nu \hat{\xi}_\mu  \right) + \partial_\mu \hat{\xi}^\alpha h_{\alpha \nu} + \partial_\nu \hat{\xi}^\alpha h_{\alpha \mu} \,  \label{dfh}   \\ &+&  \hat{\xi}^\alpha \partial_\alpha  h_{\mu \nu} + \zeta   \partial_z h_{\mu \nu}  \nonumber \\
\delta F &=&   \zeta \partial_z A  + \zeta \partial_z F + \hat \xi^\rho \partial_\rho F \\
\delta G &=&  \partial_z \big [\zeta (1+G)\big]  - \zeta \partial_z A \left(1+ G \right) + \hat \xi^\mu \partial_\mu G \, \label{dfG}
\eeq
where only the variation of $G$ changes due to the coordinate transformation. The fifth dimensional shift is parametrized as $\xi^5 \to  z + \zeta $,  related to the $y$-coordinate in the following way:
\beq
\partial_5 = e^{A} \partial_z \,, \quad \epsilon = e^{-A} \zeta
\eeq
To verify that Eq.(\ref{dfh}-\ref{dfG}) are correct infinitesimal transformation, we  directly calculate the variation of $\sqrt{g}$ 
in the conformal coordinate, which yields:
\beq
\delta_{\zeta} \sqrt{g}  &=& \delta_{\zeta} \Big( e^{-5A-4F} \sqrt{\hat g} (1+G) \Big) \nonumber \\
 &=&   e^{-5A-4F} \Big[ (\delta_{\zeta} G - 4 (1+ G) \delta_{\zeta} F) \sqrt{\hat g} \nonumber \\
 &+& (1+ G) \delta_{\zeta}\sqrt{\hat g} \Big]  \label{dfg}
\eeq
Substituting Eq.(\ref{dfh}-\ref{dfG}) into Eq.(\ref{dfg}),   we obtains that:
\begin{eqnarray}
\delta_{\zeta} \sqrt{g} 
 &=& e^{-5A-4F} \Big[\partial_z \left(\zeta (1+G)\right)- 5 \partial_z A \zeta (1+G)  \Big] \sqrt{\hat g}  \nonumber \\
&+ &  e^{-5A-4F} (1+G) \zeta \left[  -4  \partial_z F \sqrt{\hat g} +   \partial_z \sqrt{\hat g} \right]  \nonumber \\
&=& \partial_z (\zeta \sqrt{g}) 
\end{eqnarray}
As anticipated,  the transformations of metric fields  precisely recover  Eq.(\ref{dg}). Note that  Eq.(\ref{hrule}-\ref{Grule}) in the $y$-coordinate also pass this simple verification:
\beq
\delta_{\epsilon} \sqrt{g}  &=& \delta_{\epsilon} \Big( e^{-4A-4F} \sqrt{\hat g} (1+G) \Big) \nonumber \\
 &=& e^{-4A-4F} \left[ \partial_5 \left(\epsilon(1+G)\right) - 4 A' \epsilon (1+G)  \right] \sqrt{\hat g}  \nonumber \\
&+& e^{-4A-4F}  (1+G) \epsilon \,\left[  - 4  F' \sqrt{\hat g} + \partial_5 \sqrt{\hat g}  \right]  \nonumber \\
&=& \partial_5 (\epsilon \sqrt{g}) 
\end{eqnarray}

\section{$n=1$ recursive relation for potential terms} \label{app:n=1}
 We can expand the potential terms in the  bulk Lagrangian up to the quadratic order, which  can be split into $\mathcal{\hat L}_{pot}^{(2)}  =   \mathcal{\hat L}_{pot, h^2}^{(2)} + \mathcal{\hat L}_{pot, rad}^{(2)} +   \mathcal{\hat L}_{pot, mix}^{(2)}  $:
\beq
\mathcal{L}_{\hat pot, h^2}^{(2)} &=& -\frac{e^{-4A}}{8 \kappa^2} \left[ \partial_5 h_{\mu \nu} \partial_5 h^{\mu\nu} - \left(\partial_5 h \right)^2 \right] -\frac{1}{2 \kappa^2} \left[ -\partial_5 \left[e^{-4A} \left( h_{\mu \nu} \partial_5 h^{\mu \nu} - \frac{1}{2} h \partial_5 h\right) \right] \right. \nonumber \\& + & \left. \frac{1}{2} \partial_5 \left[ e^{-4A} A' \left(h_{\mu \nu} h^{\mu \nu} - \frac{1}{2} h^2 \right) \right]+ \frac{1}{2} e^{-4A} \left[ h^{\mu \nu} h_{\mu \nu} -\frac{1}{2} h^2 \right] \left[ 3 A'' - \kappa^2 \phi^{\prime 2}_0 \right] \right]  \label{eq:hkin}
\eeq
\beq
\mathcal{\hat L}_{pot, rad}^{(2)}&=&  \frac{6}{\kappa^2}  e^{-4A} \Big[  F'^2 -2  F'A' G  + G^2 A'^2  + 4 F^2 A'' \Big]   - \frac{1}{2} e^{-4A}\left[   \varphi'^2+   \left[ G^2 + 16 F^2\right]  \phi'^2_0  \right. \nonumber  \\ &-&  \left. 2 (G+ 4F) \phi'_0 \varphi'  +\left[ 2 (G- 4F)  \frac{\partial V}{\partial \phi_0} \varphi  +   \frac{\partial^2 V}{\partial \phi_0^2} \varphi^2 \right] \right] \nonumber \\ &-&\frac{4}{\kappa^2} \partial_5 \left[ e^{-4A} \left[ (G+ 4G) F'  - A' (2 F^2 + G^2 ) - 4 A' FG\right]\right]  \\ \nonumber \\
 \mathcal{\hat L}_{pot, mix}^{(2)} &=& -\frac{3}{2 \kappa^2}  e^{-4A}  \left( F' -A' G -\frac{\kappa^2}{3} \phi'_0 \varphi \right) h'  - \frac{1}{2} \partial_5 \left( e^{-4A} h \phi'_0 \varphi \right) \nonumber \\ &+& \frac{1}{2 \kappa^2} \partial_5 \left[ e^{-4A} G \left(h'  -4 A' h \right) \right] + \frac{2}{\kappa^2} \partial_5 \left[ e^{-4A} \left[ \partial_5(F h ) - A' Fh \right]\right]  \eeq
where we keep all the surface terms in addition to the normal terms. The last term in the second line of  Eq.(\ref{eq:hkin}) is a boundary term that can be cancelled by an identical term from $S_{\rm brane}$. 
\begin{itemize}
\item[(1)] For the $\epsilon$ variation of  graviton potential terms, we can derive that:
\beq
\delta^{(2)}_{\epsilon} \mathcal{\hat L}_{pot, h}^{(1)} &=& -\frac{1}{ 2 \kappa^2} \partial_5  \left[e^{-4A} \left(\partial_5 (\epsilon h') - A' \epsilon h'  \right)\right]  - e^{-4A}  \epsilon h' \left[ 3A'' -\kappa^2 \phi_0^{\prime 2}\right] \label{eq:pot1}
\eeq
\beq
\delta^{(1)}_{\epsilon} \mathcal{\hat L}_{pot, mix}^{(2)} &=& -\frac{3}{ 2\kappa^2}    e^{-4A}  \left[  \left(A' \epsilon \right)' -\epsilon' A' -\frac{\kappa^2}{3}\epsilon \phi^{\prime 2}_0  \right] h' -\frac{1}{2} \partial_5 \left( e^{-4A} \epsilon  \phi^{\prime 2}_0  h \right)
 \nonumber \\  &+&\frac{1}{2\kappa^2} \partial_5 \left[ e^{-4A} \epsilon'  \left(h'  -4 A' h \right)\right] + \frac{2}{\kappa^2} \partial_5 \left[ e^{-4A} \left[ \partial_5(A' \epsilon h ) - A'^2 \epsilon h \right]\right] \label{eq:pot2}
\eeq
Add Eqs.(\ref{eq:pot1}-\ref{eq:pot2}) up,  we obatin:
\beq
\delta^{(2)}_{\epsilon} \mathcal{\hat L}_{pot, h}^{(1)} + \delta^{(1)}_{\epsilon} \mathcal{\hat L}_{pot, mix}^{(2)} &=& -\frac{1}{2\kappa^2} \partial_5  \Big[ \epsilon \Big( \partial_5 \left[ e^{-4A}\left( h' - A' h \right)\right] -  e^{-4 A} h \left[ 3  A'' - \kappa^2   \phi^{\prime 2}_0 \right] \Big) \Big]  \nonumber \\
&=& \partial_5 \left[ \epsilon  \mathcal{\hat L}_{pot, h}^{(1)}\right] \label{eq:dpot1}
\eeq

\item[(2)]  Then we consider the variation of  gravition term with respect to $\hat \xi^\mu$:
\beq
\delta^{(2)}_{\xi} \mathcal{\hat L}_{pot, h}^{(1)} &=& -  \frac{1}{2\kappa^2} \Big\{ \partial_5 \left[ e^{-4A} \left[\partial_5 \left(2 \partial_\mu \hat \xi_\alpha  h^{\mu \alpha}  + \hat \xi^\mu \partial_\mu h \right)-A'  \left(2 \partial_\mu \hat \xi_\alpha  h^{\mu \alpha}  + \hat \xi^\mu \partial_\mu h \right) \right] \right]  \nonumber \\ 
&& - \, e^{-4A} \left(2 \partial_\mu \hat \xi_\alpha  h^{\mu \alpha}  + \hat \xi^\mu \partial_\mu h \right) \left[ 3  A'' - \kappa^2   \phi^{\prime 2}_0 \right] \Big\} \label{eq:pot3}
\eeq

\beq
\delta^{(1)}_{\xi} \mathcal{\hat L}_{pot, h^2}^{(2)} &=& -\frac{1}{2\kappa^2}\Big\{ \partial_5 \left[ e^{-4A} \left( \partial_\mu \hat \xi^\mu  h' - 2 \partial_\mu \hat \xi_\nu \partial_5 h^{\mu \nu }\right) \right]+\partial_5 \left[ e^{-4A} A' \left( 2 \partial_\mu \hat \xi_\nu h^{\mu \nu} -\partial_\mu \hat \xi^\mu h \right) \right] \nonumber \\ &+& e^{-4A} \left(2 \partial_\mu \hat \xi_\nu h^{\mu \nu} - \partial_\mu \hat \xi^\mu h \right) \left[ 3 A'' - \kappa^2 \phi^{\prime 2}_0 \right] \Big\} \label{eq:pot4}
\eeq
The summation of  Eqs.(\ref{eq:pot3}-\ref{eq:pot4}) gives:
\beq
\delta^{(2)}_{\xi} \mathcal{\hat L}_{pot, h}^{(1)} + \delta^{(1)}_{\xi} \mathcal{\hat L}_{pot, h^2}^{(2)} &=& -\frac{1}{2\kappa^2} \partial_\mu \left[ \hat \xi^\mu \Big( \partial_5 \left[ e^{-4A} \left(h'- A' h \right)\right] -e^{-4A} h \left[ 3 A'' -\kappa^2 \phi^{\prime 2}_0\right]\Big) \right] \nonumber \\
&=& \partial_\mu \left[ \hat \xi^\mu  \mathcal{\hat L}_{pot, h}^{(1)}\right] \label{eq:dpot2}
\eeq
\end{itemize}
Note that in Eq.(\ref{eq:dpot1}-\ref{eq:dpot2}), the boundary term $\propto \left[ 3  A'' - \kappa^2   \phi^{\prime 2}_0 \right] $ from the bulk Lagrangian expansion is naturally included. It is also straightforward to verify that:
\beq
 \delta_\epsilon^{(2)}  \mathcal{L}^{(1)}_{pot ,rad} +  \delta^{(1)}_\epsilon \mathcal{L}_{pot, rad}^{(2)} & = & \partial_\epsilon \left[\epsilon  \mathcal{L}_{pot, rad}^{(1)}\right]   \label{eq:dpot3}\\
  \delta_\xi^{(2)}  \mathcal{L}^{(1)}_{pot, rad} +  \delta^{(1)}_\xi \mathcal{L}_{pot, mix}^{(2)}  &=& \partial_\mu \left[\hat \xi^\mu  \mathcal{L}_{pot, rad}^{(1)}\right] \label{eq:dpot4}
\eeq
 Therefore combining Eq.(\ref{eq:dpot1}-\ref{eq:dpot2}) and Eq.(\ref{eq:dpot3}-\ref{eq:dpot4}), we  explicitly proved that:
\beq  
\delta^{(2)} \mathcal{\hat L}_{pot}^{(1)} +  \delta^{(1)} \mathcal{\hat L}_{pot}^{(2)}  =  \partial_M \left( \xi^M  \mathcal{\hat L}_{pot}^{(1)} \right) 
\eeq

\section{Derivation for linear expansion in  Eq.(\ref{eq:Lkin(1)}-\ref{eq:Lpot(1)})}\label{app:tad}

First of all, we derive the linear graviton term from the bulk Lagrangian expansion. Using the linear expansion for $R_{MN}$ in Appendix of  Ref.\cite{Cai:2021mrw}, and applying the background (BG) equation: $V(\phi_0) = -\frac{6}{\kappa^2} A'^2  + \frac{1}{2} \phi^{\prime 2}_0$, one can obtain:
\beq
&&  -\frac{1}{2 \kappa^2} \int d^5 \sqrt{g}  \left( R +  2 \kappa^2 V(\phi) \right) + \frac{1}{2} \int d^5 x \sqrt{g}  g^{IJ} \partial_I \phi \partial_J \phi  \nonumber \\  \supset
&&   -\frac{1}{2 \kappa^2} \int d^5 x  e^{-4 A} \left [ e^{2A} (\partial_{\mu} \partial_\nu h^{\mu \nu}  -\Box h) + \frac{1}{2} h'' - 4A' h' + 4 h \left( A'^2 - A''\right) + \frac{1}{2} (h'' - 2A' h') + \kappa^2 h  \phi^{\prime 2}_0\right] \nonumber \\
= &&  -\frac{1}{2 \kappa^2} \int d^5 x  \left\{ e^{-2A} (\partial_{\mu} \partial_\nu h^{\mu \nu}  -\Box h)  +  \partial_5 \left[ e^{-4A} \left( h' -  A' h \right) \right] -  e^{-4 A}  h \left[ 3  A'' - \kappa^2   \phi^{\prime 2}_0 \right] \right\} \label{tad1}
\eeq
where all three terms  match the linear graviton terms  in Eq.(\ref{eq:Lkin(1)}-\ref{eq:Lpot(1)}). In particular the last term exactly cancels the  $h$ term  from the brane action  $S_{\rm branes} = - \int d^5x \sqrt{g_4} \sum_{i} \lambda_{i}(\phi) \delta(y - y_i) $, because of  $A'' = \frac{\kappa^2}{3} \phi^{\prime 2}_0  + \frac{\kappa^2}{3} \sum_i  \lambda_i (\phi_0) \delta(y-y_0) $. Therefore only the first two total derivative terms survive in the 5d action $S= S_{\rm bulk}+ S_{\rm brane}$.

Then for the linear radion terms,  we remove the brane contribution in Eq.(A10) of Ref.\cite{Cai:2021mrw}, and affix the kinetic term to obtain: 
\beq
\mathcal{L}_{tad} &= & \frac{1}{2 \kappa^2} \int d y \, 8\,  e^{-4A}  \left(  \left[F''-A' G' \right] - 2 A'' G   - 5 A' \left[ F'- A' G\right] \right) \nonumber \\
& - &  \frac{1}{2 \kappa^2} \int d y  e^{-4A} \left[ G -4 F  \right] \left[\left( 20 A'^2 - 8 A''\right) + 2 \kappa^2 V(\phi_0)\right] \nonumber \\
&- &  \frac{1}{\kappa^2} \int  dy e^{-2 A} \Box \left( 3F - G\right)  \,. \label{tad0}
\eeq
with $V(\phi_0) = -\frac{6}{\kappa^2} A'^2  + \frac{1}{2} \phi^{\prime 2}_0$. Another contribution is from  the GW scalar $\phi = \phi_0 + \varphi $, which reads: 
\beq
\tilde{\mathcal{L}}_{tad} &=&  -\int d y e^{-4A} \left[  \phi'_0 \varphi'   - \frac{1}{2} \left( G+4F \right) \phi^{\prime 2}_0 + \frac{\partial V}{\partial \phi_0} \varphi \right]  \,. \label{tad1}
\eeq
Adding up Eq.(\ref{tad0}) and Eq.(\ref{tad1}) gives:
\beq
\mathcal{L}_{tad} +  \tilde{\mathcal{L}}_{tad} & = &- \frac{1}{\kappa^2} \int  dy e^{-2 A} \Box \left( 3F - G\right)    \nonumber \\
  &=& - \frac{1}{2 \kappa^2}  \int dy \left\{ e^{-4A} 8 A' \left( F' -A' G \right) - 8\partial_5 \left[   e^{-4A} (F'- A'G)\right]\right\} \nonumber \\
&-& \frac{1}{2 \kappa^2} \int dy e^{-4A} \left[ 8 (G- 4F) A'^2 + 32 A'' F\right] \nonumber \\
&+& \int dy e^{-4A} \left[ 4 F \phi^{\prime 2}_0 - \phi'_0 \varphi'  - \frac{\partial V}{\partial \phi_0} \varphi\right] 
\eeq
where we use the identity  $e^{-4A} \left[\left(F'' -A' G' - A'' G\right) -4A'(F'-4A'G)\right]=\partial_5 \left[ e^{-4A} \left(F'-A'G \right) \right]$ to simplify the first line in Eq.(\ref{tad0}). The $F \phi^{\prime 2}_0$ term can be removed by the BG equation: $A'' = \frac{\kappa^2}{3} \phi^{\prime 2}_0  + \frac{\kappa^2}{3} \sum_i  \lambda_i (\phi_0) \delta(y-y_0) $. Then the remaining $A''$ terms fit into a total derivative $8 e^{-4A}\left(F'A'-4FA'^2 + FA'' \right)= 8 \partial_5 \left( e^{-4A} FA' \right)$, and the  radion tadpoles  turn out to be:
\beq
\mathcal{L}_{tad} +  \tilde{\mathcal{L}}_{tad}  &=& -\frac{1}{\kappa^2} \int  dy e^{-2 A} \Box \left( 3F - G\right) +  \frac{1}{2 \kappa^2}  \int dy \left\{  8 \partial_5 \left[   e^{-4A} (F'- A'G- A' F)\right] \right\} \nonumber \\
&+& \int dy e^{-4A} \left[ - \phi'_0 \varphi'  - \frac{\partial V}{\partial \phi_0} \varphi\right] -  \int dy e^{-4A} 4 F\sum_i \lambda_i \delta(y-y_i)
\eeq
Now we can apply another BG equation: $\phi_0 '' = 4 A' \phi_0 ' + \frac{\partial V(\phi_0)}{\partial \phi}+ 
\sum_{i} \frac{\partial \lambda_i(\phi_0)}{\partial \phi} \delta(y-y_i)$  to get:
\beq
\mathcal{L}_{tad} +  \tilde{\mathcal{L}}_{tad}  &=& - \frac{1}{\kappa^2} \int  dy e^{-2 A} \Box \left( 3F - G\right) + \frac{4}{ \kappa^2}  \int dy    \partial_5 \left[  e^{-4A}  (F'- A'G- A' F)\right] \nonumber \\ 
&-&  \int dy \partial_5\left[ e^{-4A} \phi'_0 \varphi \right] - \int d y e^{-4A} \sum_i \left(4F \lambda(\phi_0)- \frac{\partial \lambda_i}{\partial \phi_0} \varphi \right) \delta (y- y_0)\label{tad2}
\eeq
which are all the  radion terms in Eq.(\ref{eq:Lkin(1)}-\ref{eq:Lpot(1)}) and the last  boundary term can be precisely cancelled by  the brane Lagrangian similar to the graviton case. Note that  we do use any linearized EOM in all the derivation,  since diffeomorphism is  an off-shell symmetry and only BG equations are applied here.

Therefore our results show that  the linear kinetic and potential terms are  total derivatives  in the 5d action  $S= S_{\rm bulk}+ S_{\rm brane}$.  This renders  the quadratic  order effective Lagrangian  by itself invariant under the linear diffeomorphism  after 5d integration.

\end{document}